\pdfoutput=1

\documentclass[11pt,twoside,a4paper,cmspaper,final,collab]{cms-tdr}
\def\svnVersion{426227P}\def\cmsCernNoTag{CERN-EP-2017-098}\def\cmsCernDate{\today}\def\cmsMessage{Published in Physical Review Letters as \href{http://dx.doi.org/10.1103/PhysRevLett.119.111802}{\doi{10.1103/PhysRevLett.119.111802.}}}

\begin{document}\cmsNoteHeader{EXO-16-030}

\hyphenation{had-ron-i-za-tion}
\hyphenation{cal-or-i-me-ter}
\hyphenation{de-vices}
\RCS$Revision: 426141 $
\RCS$HeadURL: svn+ssh://svn.cern.ch/reps/tdr2/papers/EXO-16-030/trunk/EXO-16-030.tex $
\RCS$Id: EXO-16-030.tex 426141 2017-09-18 14:21:28Z ntran $
\newcommand{\gDM}{\ensuremath{g_\mathrm{DM}}\xspace}
\newcommand{\gq}{\ensuremath{g_{\PQq}}\xspace}
\newlength\cmsFigWidth
\ifthenelse{\boolean{cms@external}}{\setlength\cmsFigWidth{0.85\columnwidth}}{\setlength\cmsFigWidth{0.4\textwidth}}
\ifthenelse{\boolean{cms@external}}{\providecommand{\cmsLeft}{upper\xspace}}{\providecommand{\cmsLeft}{left\xspace}}
\ifthenelse{\boolean{cms@external}}{\providecommand{\cmsRight}{lower\xspace}}{\providecommand{\cmsRight}{right\xspace}}
\ifthenelse{\boolean{cms@external}}{\providecommand{\xtolerance}{\relax}}{\providecommand{\xtolerance}{\tolerance=1200}}
\providecommand{\NA}{\ensuremath{\text{---}}\xspace}
\providecommand{\CLs}{\ensuremath{\mathrm{CL}_\mathrm{s}}\xspace}
\ifthenelse{\boolean{cms@external}}{\providecommand{\CL}{C.L.\xspace}}{\providecommand{\CL}{CL\xspace}}

\cmsNoteHeader{EXO-16-030}
\title{Search for low mass vector resonances decaying to quark-antiquark pairs in proton-proton collisions at \texorpdfstring{$\sqrt{s}=13\TeV$}{sqrt(s) = 13 TeV}}

\date{\today}

\abstract{
A search is reported for a narrow vector resonance decaying to quark-antiquark pairs in proton-proton collisions at $\sqrt{s}=13\TeV$, collected with the
CMS detector at the LHC.
The data sample corresponds to an integrated luminosity of 2.7\fbinv.
The vector resonance is produced at large transverse momenta,
with its decay products merged into a single jet. The resulting signature is a peak over background
in the distribution of the invariant mass of the jet.
The results are interpreted in the framework of a leptophobic vector resonance and
no evidence is found for such particles in the mass range of 100--300\GeV.
Upper limits at 95\% confidence level on the production cross section are presented in a region of mass-coupling phase space
previously unexplored at the LHC. The region below 140\GeV has not been explored by any previous experiments.
}

\hypersetup{%
pdfauthor={CMS Collaboration},%
pdftitle={Search for low mass vector resonances decaying to quark-antiquark pairs in proton-proton collisions at sqrt(s) = 13 TeV},%
pdfsubject={CMS},%
pdfkeywords={CMS, physics, vector resonances}}

\maketitle

Over the past half-century, searches for narrow resonances in the dijet invariant mass spectrum have been an important part of the physics program at every collider.
Such searches are well motivated within the many classes of theories beyond the current standard model (SM) that predict resonances with significant couplings to quarks~\cite{ref_gauge,ref_axi,ref_qstar,ref_diquark,Baur:1989kv,ref_coloron,ref_rsg,Cullen:2000ef,Anchordoqui:2008di,Han:2010rf,Chivukula:2011ng}.
The first searches for such particles at the UA1~\cite{Albajar:1988rs} and
UA2~\cite{Alitti:1990kw,Alitti:1993pn} experiments
at the CERN S$\PAp\Pp$S have been extended to larger
values of resonance masses by the CDF~\cite{Abe:1989gz,Abe:1993it,Abe:1995jz,Abe:1997hm,Aaltonen:2008dn}
and D0~\cite{D0-1,D0-2,Abazov:2003tj}
experiments at the Fermilab Tevatron, and by
the ATLAS~\cite{ATLAS1,ATLAS2,ATLAS3,ATLAS4,ATLAS5,ATLAS6,ATLAS7,ATLAS8,ATLAS9,Aaboud:2017yvp}
and CMS~\cite{CMS1,CMS2,CMS3,CMS4,CMS5,CMS6,CMS7,CMS8,CMS9,CMS10,CMS11,Sirunyan:2017ygf,Sirunyan:2016iap}
experiments at the CERN LHC.

In this Letter, we report on a search for
vector (\PZpr) resonances decaying to quark-antiquark pairs ($\qqbar$) using the data collected in pp collisions at $\sqrt{s} = 13$\TeV by the
CMS detector, corresponding to an integrated luminosity of
2.7\fbinv.  This search concentrates on the mass region of
100--300\GeV, exploring for the first time masses below
140\GeV~\cite{Alitti:1990kw,Alitti:1993pn}.
To access this mass regime, we present a new analysis technique exploiting novel jet substructure techniques 
and associated production with a jet. 
The search is interpreted in the framework of a leptophobic vector resonance and can also be reinterpreted in terms of searches for generic vector-like resonances decaying to quarks.~\cite{Dobrescu:2013coa}.

With the increase in collision energy and beam intensity at hadron
colliders, there has been a loss of search sensitivity for lighter resonances
with couplings to quarks and gluons.  The main experimental difficulties
originate from the large increase in the cross section of multijet backgrounds at small resonance masses, and the more restrictive
trigger requirements needed to reduce the data recording rate because of limited resources for
event processing and storage.
To overcome these
difficulties, we require events to have a jet with large transverse momentum (\pt) from
initial-state radiation (ISR) produced in association with the
resonance.  This ISR constraint provides enough energy in the event to satisfy the trigger.
Combinatorial backgrounds are reduced by requiring the resonance be reconstructed within a single jet.
The jet is required to have the two-prong structure expected from signal.
The dominant background
is from SM events comprised of jets produced through quantum chromodynamics (QCD).
This multijet background is estimated by
inverting the two-prong substructure requirement, which is specifically designed to be uncorrelated with the jet mass.
By searching for new particles produced in association with an ISR jet, we are able to
search for new resonances in a coupling and mass regime to which previous searches were insensitive.

A detailed description of the CMS
detector, together with a definition of the coordinate system and
the relevant kinematic variables, can be found
in Ref.~\cite{Chatrchyan:2008zzk}.
The central feature of the CMS apparatus is a superconducting solenoid
of 6\unit{m} internal diameter, providing a magnetic field of
3.8\unit{T}. A silicon
pixel and strip tracker, a lead tungstate crystal electromagnetic
calorimeter (ECAL), and a brass and scintillator hadron calorimeter
(HCAL), each composed of a barrel and two endcap sections, reside within the solenoid. Forward
calorimeters extend the pseudorapidity ($\eta$)~\cite{Chatrchyan:2008zzk}
coverage provided by the barrel and endcap detectors. Muons are
measured in gas-ionization detectors embedded in the steel flux-return
yoke outside the solenoid.

We employ the CMS particle-flow (PF)
algorithm~\cite{PFT-09-001,CMS-PAS-PFT-10-001} to reconstruct and identify each
individual particle through an optimized combination of information from
the various elements of the CMS detector. The energy of a photon is
obtained directly from the ECAL measurement, corrected for
losses in detector sensitivity due to zero suppression near signal threshold.
The energy of an electron is determined from a
combination of its momentum at the primary interaction vertex
determined using the tracker, the energy of the corresponding ECAL
cluster, and the energy sum of all bremsstrahlung photons spatially
compatible with originating from the electron track. The energy of a
muon is obtained from the curvature of the corresponding track. The
energy of a charged hadron is obtained from a combination of its
momentum measured in the tracker and the matching of ECAL and HCAL energy
depositions, corrected for the response
function of the calorimeters to hadronic showers. The energy
of a neutral hadron is obtained from the corresponding corrected ECAL
and HCAL energies.

{\xtolerance
The benchmark \PZpr signal events are simulated at leading order (LO) using the {\MGvATNLO} 2.2.1 generator~\cite{Alwall:2014hca}; {\sc Pythia} 8.205 generator~\cite{Sjostrand:2006za,Sjostrand:2007gs} and  {\GEANTfour}~\cite{Agostinelli:2002hh}. The dominant SM backgrounds arise from multijets,
$\PW \to \qqbar$, and $\PZ \to \qqbar$ sources.
These backgrounds are simulated at LO using the {\MGvATNLO} 2.2.1 generator and hadronized with {\PYTHIA} using the CUETP8M1 tune~\cite{Khachatryan:2015pea}.
To events containing \PW~and \PZ\ bosons, we apply higher-order QCD and Electroweak (EWK) corrections to improve modeling of the
high \pt \PW~and \PZ\ events, following~\cite{Sirunyan:2017hci,Kallweit:2014xda,Kallweit:2015dum,Kallweit:2015fta,Schonherr:2016pvn}.
Owing to the similarity between the signal process and the SM \PW~and \PZ\ topologies, the next-to-leading-order (NLO) QCD corrections and associated uncertainties applied to the \PW~and \PZ\ simulations are also applied to the signal simulation.
The NLO EWK corrections are not applied since such corrections are different for leptophobic \PZpr signal production.
\par
}

To isolate the \PZpr signal and overcome trigger restrictions, we
select a high-\pt signal jet, which typically recoils against
another high-\pt ISR jet.  We use a logical ``or" of trigger
requirements that selects on the total hadronic transverse
energy in the event and, in some cases, additionally select on the mass of the jet after removing remnants of soft radiation,
with the jet trimming technique~\cite{Krohn:2009th}.  Jets at the
trigger level are reconstructed using the anti-\kt
algorithm~\cite{Cacciari:2008gp,Cacciari:2011ma} with a distance parameter of 0.8, and are referred to as AK8 jets.

After the trigger selection, the AK8 jets are reconstructed by clustering particle candidates in the event.
To mitigate the effect of additional interactions in the same or adjacent bunch crossings (pileup), the pileup per particle identification (PUPPI)
algorithm~\cite{Bertolini:2014bba} is used, prior to jet clustering, to weight the PF
candidates on the likelihood of coming from the primary interaction vertex.  Other corrections are applied to jet
energies as a function of jet $\eta$ and \pt to match the detector response and to bring data and simulation into agreement.
We require at least one AK8 jet
to have $\pt > 500\GeV$ and to satisfy $\abs{\eta} < 2.5$ to be fully efficient with respect to the trigger requirements.  We veto
events containing identified and isolated electrons or muons with
$\pt > 10\GeV$, and $\abs{\eta} < 2.5$ or $2.4$, respectively, to reduce backgrounds from SM EWK processes.

The Lorentz-boosted $\PZpr \to \qqbar$ system is reconstructed as a single high-\pt jet,
where the decay products have merged into one object, assumed to be the jet of highest \pt in the event.
From simulation studies, approximately 70\% of \PZpr signal events satisfy this assumption.
The most important variable is the soft drop jet mass ($m_\mathrm{SD}$), which employs soft drop grooming~\cite{Dasgupta:2013ihk,Larkoski:2014wba}, a technique used to remove soft and
wide-angle radiation in jets.
The variable $m_\mathrm{SD}$ enhances the peak at the \PZpr mass for signal
events and the same act of soft drop grooming reduces the masses from background quark- and gluon-initiated
events. For computing $m_\mathrm{SD}$, we use a soft
radiation fraction selection greater than $z= 0.1$ and an angular exponent
parameter of $\beta = 0$.

Since the $\PZpr \to
\qqbar$ jet has a two-prong structure, we use substructure
tools developed to identify jets from hadronically decaying
\PW~bosons. We use a variable based on the $N$-subjettiness
ratio $\tau_2/\tau_1$~\cite{Thaler:2010tr}, also referred to as $\tau_{21}$, which is used in other
searches
to determine a jet's consistency with having a two-prong
structure~\cite{Khachatryan:2014vla,Khachatryan:2014gha,Khachatryan:2014hpa}.  However, any chosen cutoff $\tau_{21}$ value sculpts the jet mass distributions in a way that depends on the \pt of the jet, and causes a complex mass distribution that peaks at high jet mass.
This limits its usefulness when searching for resonant peaks over a large range in \pt.
We therefore employ a transformation of
$\tau_{21}$ to $\tau_{21}^\mathrm{DDT}$, where DDT stands for designed decorrelated tagger~\cite{Dolen:2016kst}.
It is a generic technique to reduce mass correlations of the observable in multijet events.
We specifically deploy it here through a linear transformation between $\tau_{21}$ and a dimensionless jet mass variable~\cite{Salam:2009jx}, $\rho^\mathrm{DDT} = \ln [m_\mathrm{SD}^2/ (\mu_{0} \pt)]$, where the scale $\mu_{0} = 1\GeV$.
We determine the optimal transformation from simulation to be
$\tau_{21}^\mathrm{DDT} = \tau_{21} + 0.063 \, \rho^\mathrm{DDT}$.
Events are selected by requiring $\tau_{21}^\mathrm{DDT} < 0.38$.

\begin{figure}[ht]
\centering
   \includegraphics[width=0.48\textwidth] {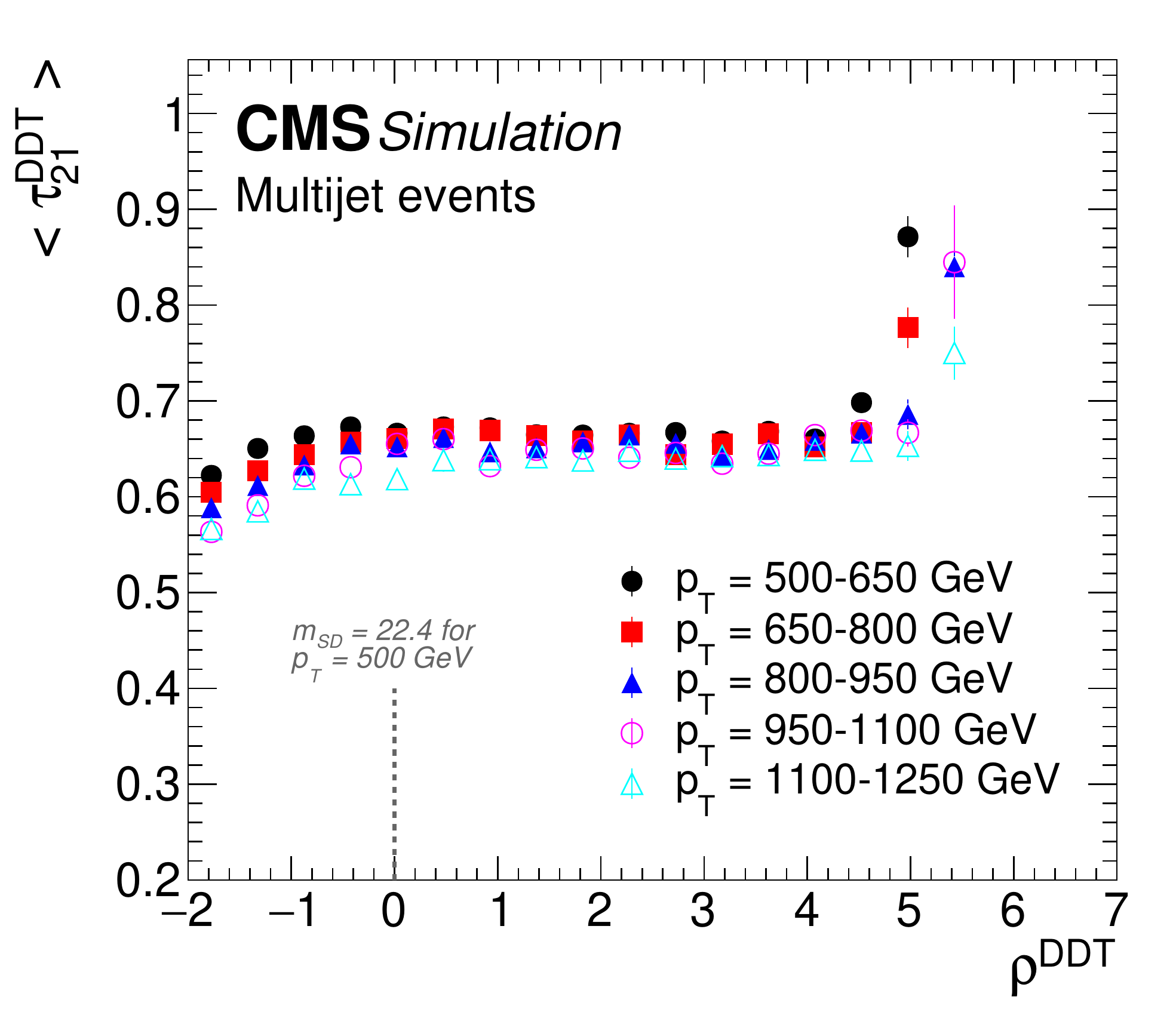}
   \includegraphics[width=0.48\textwidth] {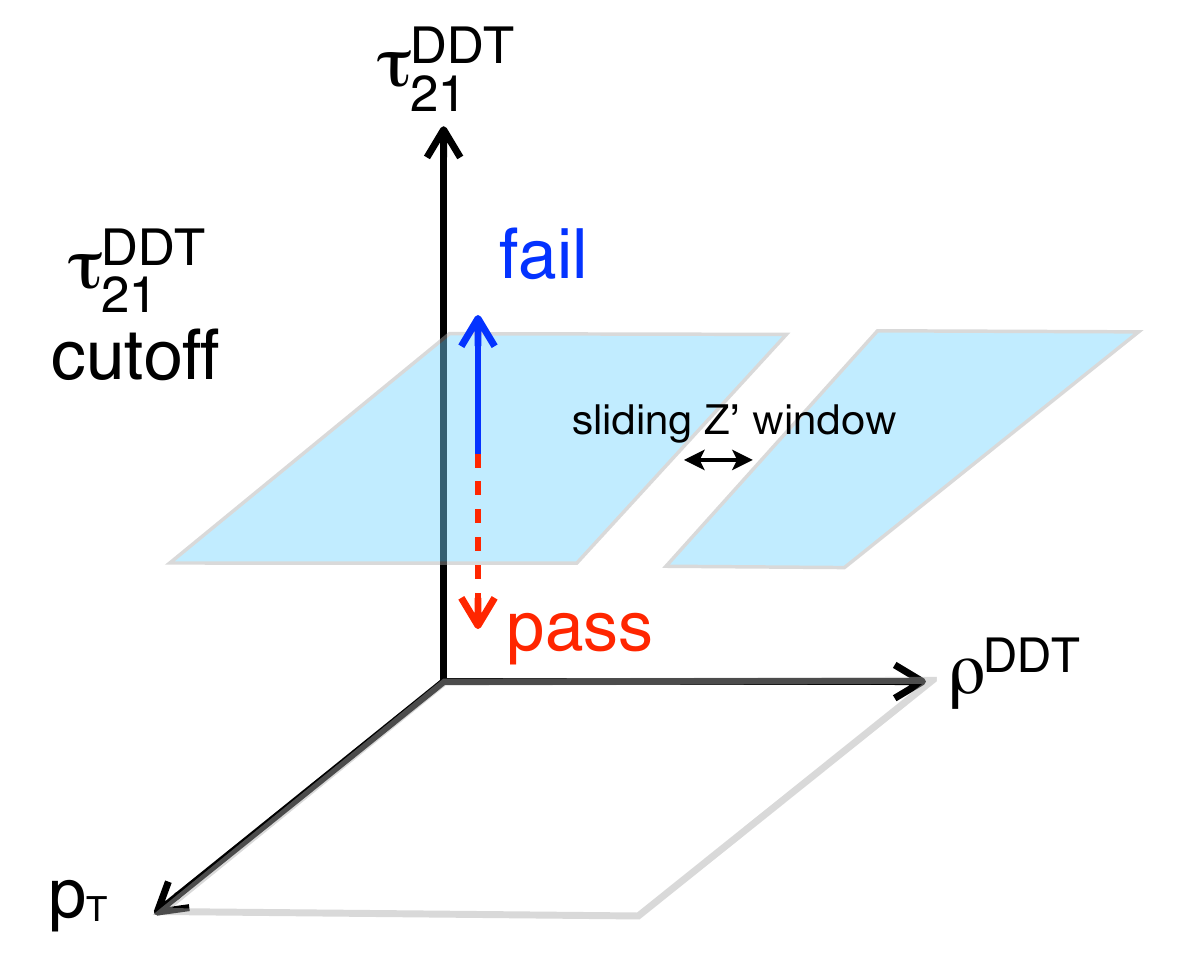}
\caption{
The average of $\tau_{21}^\mathrm{DDT}$ versus $\rho^\mathrm{DDT}$ for different bins in jet \pt (\cmsLeft). A schematic of the method used to estimate the background (\cmsRight).
After a $\tau_{21}^\mathrm{DDT}$ selection is applied, the multijet event distributions in the passing region are predicted from their distributions in the failing region with the use of a pass-to-fail ratio extracted through a 2-dimensional fit in $( \rho^\mathrm{DDT}, \pt )$ space.
Events in a sliding window around the desired \PZpr mass point are vetoed when determining the pass-to-fail ratio.
\label{fig:rhalphstrat2}
}
\end{figure}

The dominant background comes from multijet events.  For each \PZpr mass point, we
estimate this background using sidebands of kinematic
distributions in the data, in order to minimize reliance on the simulation.
We use both $\rho^\mathrm{DDT}$ and $\tau_{21}^\mathrm{DDT}$ background-dominated sideband regions and
take advantage of the lack of correlation between
$\tau_{21}^\mathrm{DDT}$ and $\rho^\mathrm{DDT}$.
The method is illustrated in
Fig.~\ref{fig:rhalphstrat2}(\cmsLeft), where the mean in the distribution of
$\tau_{21}^\mathrm{DDT}$ is plotted as a function of $\rho^\mathrm{DDT}$ for different
bins in jet \pt.  Both $\tau_{21}^\mathrm{DDT}$ and \pt have little dependence on $\rho^\mathrm{DDT}$ in range from 0 to 4 corresponding to $30\GeV < m_\mathrm{SD} < 300\GeV$.
Figure~\ref{fig:rhalphstrat2}(\cmsRight) illustrates the strategy for estimating the multijet background.
We define the pass-to-fail ratio, $R_{p/f}$, as the ratio of yields in the pass region ($\tau_{21}^\mathrm{DDT} < 0.38$) to the yields in the fail region ($\tau_{21}^\mathrm{DDT} > 0.38$).
We rescale multijet events from the fail region using $R_{p/f}$ parameterized in \pt and $\rho^\mathrm{DDT}$ to predict the mulijet signal region background:
\begin{linenomath}
\begin{equation}
P_\text{pass}^\text{multijet}(\pt,\rho^\mathrm{DDT}) = R_{p/f}(\pt,\rho^\mathrm{DDT}) \, P_\text{fail}^\text{multijet}(\pt,\rho^\mathrm{DDT})
\end{equation}
\end{linenomath}
where $P_\text{pass}^\text{multijet}$ and $P_\text{fail}^\text{multijet}$ are the passing and failing multijet distributions.
We determine $R_{p/f}$ by fitting the ratio of data events in the pass
to fail regions over the ($\rho^\mathrm{DDT}$,~\pt) space.  With this method, we fit for any residual correlations between
$\tau_{21}^\mathrm{DDT}$ and $\rho^\mathrm{DDT}$, modeling them as polynomial functions.  
Fitting $R_{p/f}$, in place of the normalization, allows the region where the trigger selection is not fully efficient to be exploited; therefore we begin the $R_{p/f}$ fit in this region at a jet \pt $>$ 350\GeV.
Prior to performing the fit, we
subtract the \PW~and \PZ boson contributions expected from simulation, at NLO, to estimate only nonresonant contributions
such as the multijet background.
To avoid a bias induced by a given \PZpr signal,
we exclude a window from the $R_{p/f}$ fit centered about the given \PZpr $m_\mathrm{SD}$ of width ${\pm}10\% \, m_{\PZpr}$, which corresponds approximately to the jet mass resolution.
The missing strip in the $R_{p/f}$ shape in Fig.~\ref{fig:rhalphstrat2}\,(\cmsRight) reflects this window.
Fits are therefore performed as a function of \PZpr mass to estimate the background.
The two-dimensional $R_{p/f}$ depends slightly on the chosen mass hypothesis, as
different windows in jet mass are excluded for each \PZpr mass.
Finally, the mass distribution in data is tested for the presence of signal using the  multijet prediction along with the simulated \PW~and \PZ predictions.

The uncertainties in the method described above arise
from three sources.
The first uncertainty arises from residual systematic effects in the method, as
determined from a self-consistency check in simulated events.
This systematic effect is correlated across all jet mass bins.
The second is due to the uncertainty in the $R_{p/f}$ shape from the fit
parameter uncertainties, also correlated across all jet mass bins.
The third source originates from uncorrelated statistical uncertainties in the size of the
fail sample and the statistical uncertainties in the self-consistency test.

Smaller backgrounds from resonant SM processes (\PW/\PZ+jets) are estimated from simulation, but with corrections for the mass shapes and efficiencies evaluated from data as discussed below.
The expectations for all SM background processes (multijets, \PW/\PZ+jets) are determined for jet masses from 30 to 330\GeV.
Including the region near the \PW~and \PZ~boson masses provides a constraint on the mass distribution of the \PZpr and its normalization by determining the  \PW/\PZ~+jets contribution in situ.

The systematic effects on shapes and normalizations of the \PW, \PZ, and signal distributions are correlated.
We constrain the jet mass
scale, jet mass resolution, and $\tau_{21}^\mathrm{DDT}$ selection
efficiency using a sample of merged \PW~boson jets in semileptonic
$\ttbar$ events in data.
Using the same $\tau_{21}^\mathrm{DDT}$ selection as in the analysis provides
passing and failing regions for merged \PW~boson jets in data and in simulation.
A simultaneous fit to the two samples in $m_\mathrm{SD}$ is performed to extract
the tagging efficiency of a merged \PW~boson jet, its jet mass scale, and the resolution in simulation and in data.
Scale factors relating data and simulation are then computed and
applied to the simulation.
These scale factors and their
uncertainties determine the initial distributions for the \PW, \PZ, and
signal that are further constrained in the final fit in
situ using the presence of the \PW~and \PZ peaks in the jet mass
distribution.
Finally, additional systematic uncertainties are applied to
the \PW, \PZ, and \PZpr signal yields
that are associated with higher-order corrections to the boson \pt distributions,
jet energy scale~\cite{jec}, the modeling of pileup and the integrated beam luminosity~\cite{lumi}.
A quantitative summary of the systematic effects considered is shown in Table~\ref{tab:sys}.

\begin{table}[hbtp]
\centering
 \topcaption{Summary of the systematic uncertainties for signal and background. Uncertainties due to the finite size of the simulated samples are applied only to the \PZpr signal. Electroweak uncertainties are applied only to the \PW/\PZ boson processes. Dashes (---) indicate the uncertainty does not apply.}
\begin{scotch}{ lll }
 Systematic effect & Multijet & SM $\PW/\PZ$ and \PZpr \\
 \hline
 Multijet fit parameters & 1--4\% &\NA\\
 Multijet bin-by-bin statistics & 1--20\% &\NA\\
 Multijet fit closure & 1--12\% &\NA\\
 \PW~boson tagging scale factor &\NA& 20\% \\
 NLO QCD \pt corrections &\NA& 8--12\% \\
 Jet energy/mass resolution &\NA& 10\% \\
 Simulation sample size &\NA& 1-30\% \\
 Integated luminosity &\NA& 2.7\% \\
 Pileup &\NA& $<$1\% \\
 Jet energy/mass scale &\NA& 1\% \\
\end{scotch}
 \label{tab:sys}
\end{table}

\begin{figure}[hbtp]\centering
    \includegraphics[width=0.48\textwidth]{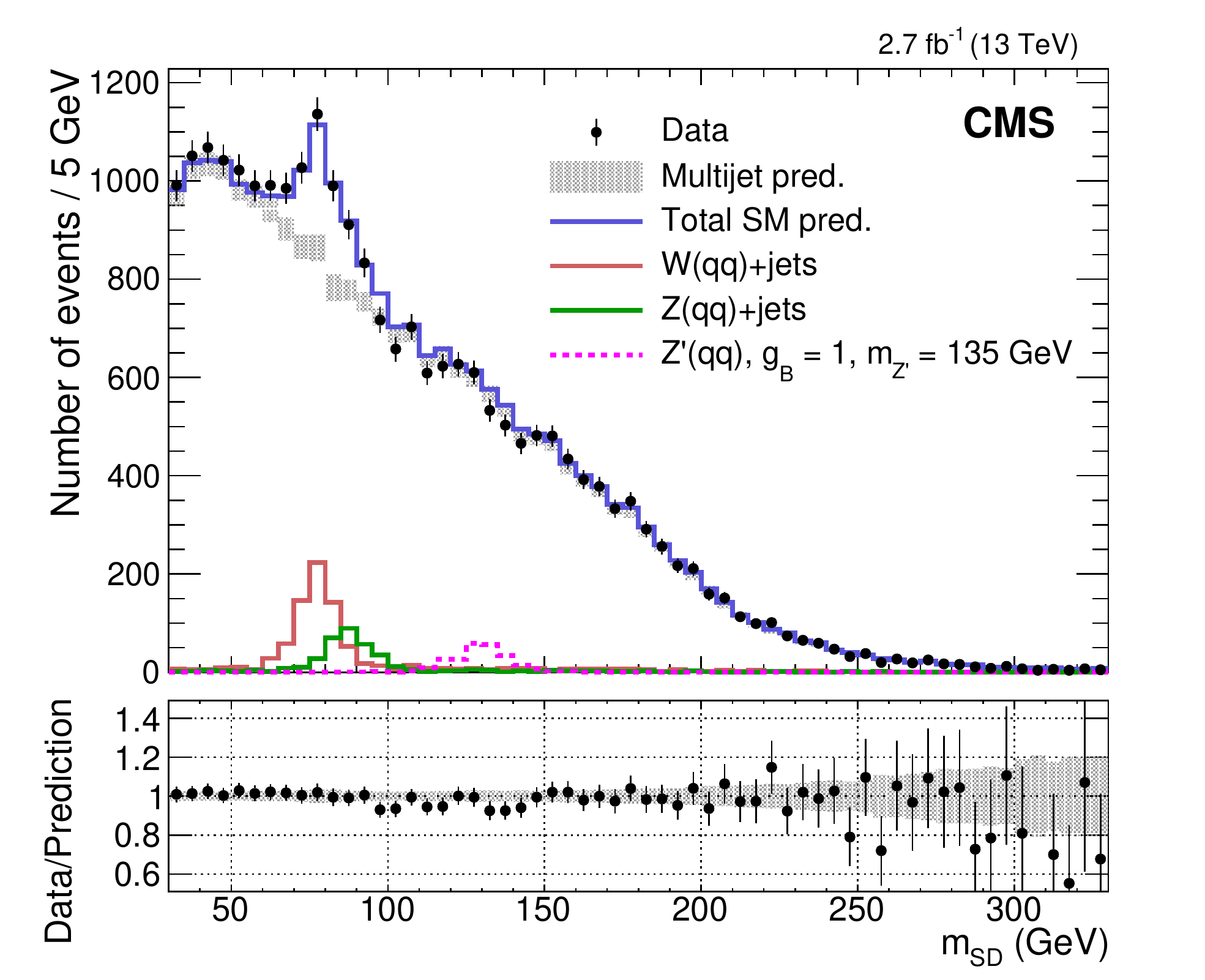}
    \caption{Soft drop jet mass distribution after final selection. Data are shown as black points.  The multijet background, including uncertainties, is shown as gray boxes, while the sum of the SM processes is given by the blue line.  Contributions from the SM \PW/\PZ boson and a hypothetical \PZpr signal at a mass of 135\GeV are also indicated.  In the bottom panel, we show the ratio of the data to the background prediction, including uncertainties. }
 \label{fig:money}
 \end{figure}

To validate the background estimation
method and its associated systematic uncertainties, studies are performed
on simulated events by injecting a signal and testing for a bias in
the extracted signal cross section.  No significant bias is observed.
We combine the estimates of various SM background processes, and
search for a \PZpr resonance in the 100--300\GeV mass range, while the SM
backgrounds are estimated in the 30--330\GeV mass range.
Figure~\ref{fig:money} shows the soft drop jet mass distribution with a background calculated for the \PZpr search at a mass of 135\GeV. Expected background contributions from the \PW~and \PZ bosons are clearly visible in the data.

The number of events in data is consistent
with the predicted backgrounds from SM processes.  Results are obtained from combined signal and background binned
likelihood fits to the jet mass distribution, such as the one shown in Fig.~\ref{fig:money}.
The results are interpreted as upper limits on the production cross
section for signal in terms of 95\% confidence level (\CL) upper limits computed using the modified frequentist
approach (\CLs), taking the likelihood as
the test statistic~\cite{CLS2,CLS1} in the asymptotic approximation~\cite{bib:asCLs}.
Systematic uncertainties are treated as nuisance
parameters that are modeled through log-normal priors.
The probability of the downward fluctuation of the data at a \PZpr signal mass of 100\GeV is less than 0.15\% 
and corresponds to a local $2.8\sigma$ effect. We attribute this to a statistical fluctuation in data. 
Studies of the background model are performed by injecting simulated signal events and we find negligible biases in the fitted signal strength.
An alternative background estimate is also studied by fitting directly the jet mass distribution, results are consistent within 1~$\sigma$.

\begin{figure}[hbtp]\centering
    \includegraphics[width=0.48\textwidth]{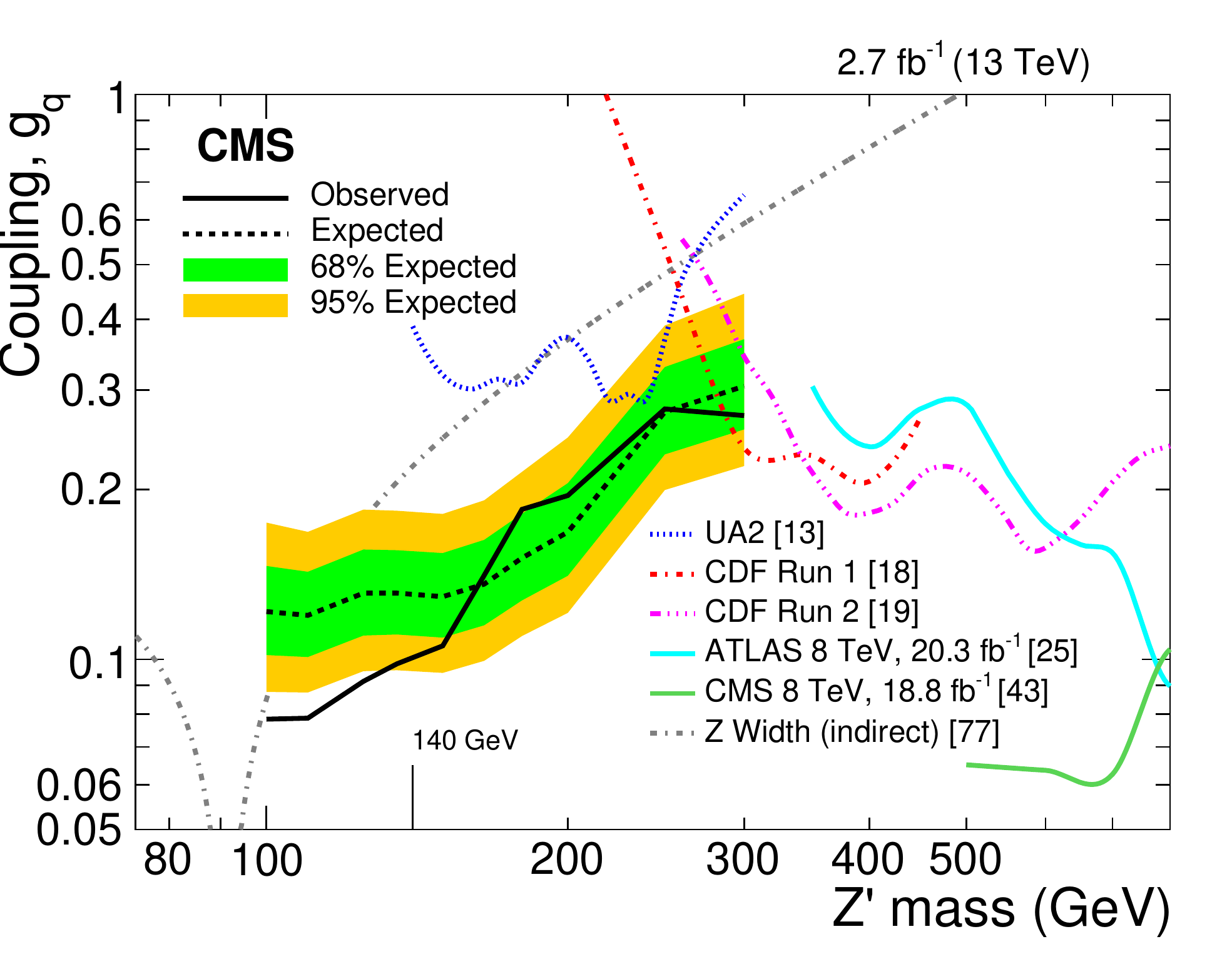}
    \caption{
	Upper limits on \gq as a function of the \PZpr mass at 95\% \CL~\cite{Boveia:2016mrp}.
    Limits from other relevant searches are shown.
    An indirect constraint from the width of the SM \PZ boson is also given.
 	\label{fig:gBlimits}
  }
  \end{figure}

Upper limits on signal cross sections are translated into the coupling of the \PZpr boson to quarks (\gq) as a function of $m_{\PZpr}$.
In Fig.~\ref{fig:gBlimits}, we compare our results with previous results from the UA2, CDF, ATLAS, and CMS experiments.  The limits  from the UA2 and CDF experiments, and the indirect constraints from the $\PZ \to \qqbar$ width are obtained following the approach described in Ref.~\cite{Dobrescu:2014fca}.  We exclude coupling values of \gq~from
0.071 to 0.260 over the selected \PZpr mass range.

In summary, a search by CMS for a vector (\PZpr) resonance decaying to a $\qqbar$ pair in the mass range from 100-300 GeV has been presented.
The results are interpreted in the framework of a leptophobic vector resonance model.
No excess above the SM prediction is observed, and 95\% confidence level upper limits are set on the
coupling of the \PZpr to quarks as a function of the \PZpr mass.  These limits are the
most stringent to date for masses less than 300\GeV.  For masses below
140\GeV, these are the only available direct limits.

\begin{acknowledgments}
We congratulate our colleagues in the CERN accelerator departments for the excellent performance of the LHC and thank the technical and administrative staffs at CERN and at other CMS institutes for their contributions to the success of the CMS effort. In addition, we gratefully acknowledge the computing centers and personnel of the Worldwide LHC Computing Grid for delivering so effectively the computing infrastructure essential to our analyses. Finally, we acknowledge the enduring support for the construction and operation of the LHC and the CMS detector provided by the following funding agencies: BMWFW and FWF (Austria); FNRS and FWO (Belgium); CNPq, CAPES, FAPERJ, and FAPESP (Brazil); MES (Bulgaria); CERN; CAS, MoST, and NSFC (China); COLCIENCIAS (Colombia); MSES and CSF (Croatia); RPF (Cyprus); SENESCYT (Ecuador); MoER, ERC IUT, and ERDF (Estonia); Academy of Finland, MEC, and HIP (Finland); CEA and CNRS/IN2P3 (France); BMBF, DFG, and HGF (Germany); GSRT (Greece); OTKA and NIH (Hungary); DAE and DST (India); IPM (Iran); SFI (Ireland); INFN (Italy); MSIP and NRF (Republic of Korea); LAS (Lithuania); MOE and UM (Malaysia); BUAP, CINVESTAV, CONACYT, LNS, SEP, and UASLP-FAI (Mexico); MBIE (New Zealand); PAEC (Pakistan); MSHE and NSC (Poland); FCT (Portugal); JINR (Dubna); MON, RosAtom, RAS, RFBR and RAEP (Russia); MESTD (Serbia); SEIDI, CPAN, PCTI and FEDER (Spain); Swiss Funding Agencies (Switzerland); MST (Taipei); ThEPCenter, IPST, STAR, and NSTDA (Thailand); TUBITAK and TAEK (Turkey); NASU and SFFR (Ukraine); STFC (United Kingdom); DOE and NSF (USA).
\end{acknowledgments}
\bibliography{auto_generated}

\cleardoublepage \appendix\section{The CMS Collaboration \label{app:collab}}\begin{sloppypar}\hyphenpenalty=5000\widowpenalty=500\clubpenalty=5000\textbf{Yerevan Physics Institute,  Yerevan,  Armenia}\\*[0pt]
A.M.~Sirunyan, A.~Tumasyan
\vskip\cmsinstskip
\textbf{Institut f\"{u}r Hochenergiephysik,  Wien,  Austria}\\*[0pt]
W.~Adam, F.~Ambrogi, E.~Asilar, T.~Bergauer, J.~Brandstetter, E.~Brondolin, M.~Dragicevic, J.~Er\"{o}, M.~Flechl, M.~Friedl, R.~Fr\"{u}hwirth\cmsAuthorMark{1}, V.M.~Ghete, J.~Grossmann, J.~Hrubec, M.~Jeitler\cmsAuthorMark{1}, A.~K\"{o}nig, N.~Krammer, I.~Kr\"{a}tschmer, D.~Liko, T.~Madlener, I.~Mikulec, E.~Pree, D.~Rabady, N.~Rad, H.~Rohringer, J.~Schieck\cmsAuthorMark{1}, R.~Sch\"{o}fbeck, M.~Spanring, D.~Spitzbart, J.~Strauss, W.~Waltenberger, J.~Wittmann, C.-E.~Wulz\cmsAuthorMark{1}, M.~Zarucki
% ...
\vskip\cmsinstskip
% ...
1:~~Also at Vienna University of Technology, Vienna, Austria\\
% ...

\end{sloppypar}
\end{document}